\documentclass [aps,twocolumn,superscriptaddress,altaffilletter,lengthcheck,tightenlines,showpacs,showkeys]{revtex4}

\newcommand{\al}{\alpha}

\newcommand{\ben}{\begin{eqnarray}}
\newcommand{\een}{\end{eqnarray}}
\newcommand{\be}{\begin{equation}}
\newcommand{\ee}{\end{equation}}
\newcommand{\ba}{\begin{eqnarray}}
\newcommand{\ea}{\end{eqnarray}}
\newcommand{\n}{\label}

\newcommand{\la}{\lambda}
\newcommand{\ga}{\gamma}
\newcommand{\ro}{\rho}

\usepackage[dvipdf]{epsfig}
\usepackage{color}
\begin{document}

%%%%%%%%%%%%%%%%%%%%%%%%%%%%%%%%%%%%%%%%%%%%%%%%%%%%%%%%%%%%%%%%%%%%%%%%%%%%%%%%%%%%%%%%%%%%%%%%%%%%%%%%%%%%%%%%%%%%%%%%%%%%%%%%%%%%%%%%

\title{k-essence and tachyons in brane-world cosmology }

%%%%%%%%%%%%%%%%%%%%%%%%%%%%%%%%%%%%%%%%%%%%%%%%%%%%%%%%%%%%%%%%%%%%%%%%%%%%%%%%%%%%%%%%%%%%%%%%%%%%%%%%%%%%%%%%%%%%%%%%%%%%%%%%%%%%%%%%%%

\author{Luis P. Chimento}\email{chimento@df.uba.ar}
\affiliation{ Departamento de F\'{\i}sica, Facultad de Ciencias Exactas y
Naturales,  Universidad de Buenos Aires, Ciudad
Universitaria, Pabell\'on I, 1428 Buenos Aires, Argentina}

\author{Mart\'{\i}n G. Richarte}\email{martin@df.uba.ar}
\affiliation{ Departamento de F\'{\i}sica, Facultad de Ciencias Exactas y
Naturales,  Universidad de Buenos Aires, Ciudad
Universitaria, Pabell\'on I, 1428 Buenos Aires, Argentina}

\begin{abstract}
We investigate a k-essence field localized on the brane evolving linearly with the cosmological time for some kinetic functions and consider the atypical k-essence model for linear and nonlinear k-fields in Friedmann-Robertson-Walker (FRW) cosmology. In the former case the k-field is driven by an inverse quadratic polynomial potential and the solutions exhibit several different behaviors which include expanding, contracting and bouncing universes as well as a model with a finite time span, some of them ending in a big crunch or a big rip. In the latter case we find the potential and show that the atypical k-essence model is dynamically uncomplete. Particularly, by selecting the extended tachyonic kinetic functions we analyze the high and low energy limits of our model, obtaining the nearly power law solution. We introduce a tachyon field with negative energy density and show that the universe evolves between two singularities.

\end{abstract}

\maketitle

%%%%%%%%%%%%%%%%%%%%%%%%%%%%%%%%%%%%%%%%%%%%%%%%%%%%%%%%%%%%%%%%%%%%%%%%%%%%%%%%%%%%%%%%%%%%%%%%%%%%%%%%%%%%
\section{Introduction}
%%%%%%%%%%%%%%%%%%%%%%%%%%%%%%%%%%%%%%%%%%%%%%%%%%%%%%%%%%%%%%%%%%%%%%%%%%%%%%%%%%%%%%%%%%%%%%%%%%%%%%%%%%%%

Recent progresses in Superstring and M-theory have offered a new perspective to understand the cosmological evolution of the universe \cite{ADD1}, \cite{HW1}, \cite{HW2}, \cite{RS1}, \cite{RS2}.
Under this scheme our Universe is conceived as a 3D brane embeded in a higher dimensional spacetime usually referred as the bulk \cite{RSH1}. One of the main features of the brane-worlds is based on the assumptions that matter fields are confined on the brane while the graviton can propagate into the bulk as well as into the brane\cite{review}. Among many interesting models coming from this new  scenarios, there are two possibilities that seem to resolve different problems in the standard cosmology. On the one hand, the RS models exhibit a set up where the extradimension could be infinite and the effective gravitational interaction posses a very small correction on the brane \cite{RS1}. On the other hand, the DGP scheme provides a modification of gravitational laws at large distances and allows, by means of a non trivial mechanism, to recover the 4D Einstein gravity at moderate scale \cite{DGP1}.

A great amount of works have been invested in studying cosmological scenarios with non-canonical kinetic term usually known as k-essence models. The idea of k-essence has its origin in Born-Infeld action of string theory \cite{BIS}. Lately,  efforts in the framework of k-essence have been directed toward model building using power law solutions which preserve \cite{CHI1}, \cite{PAD1},\cite{AFE1}, \cite{CHFE1} or violate the weak-energy condition \cite{ACHR}. In addition, a k-essence model with a divergent sound speed (called atypical k-essence) was carefully analyzed in  Refs. \cite{MCL}, \cite{CHIR1}. In the latter work it was shown that the model fixes the form of the Lagrangian for k-essence matter. Many others aspects regarding  k-essence theory have been explored in the literature  such as an unifying model of  dark components with k-essence \cite{BM1}, k-essence as dark energy source \cite{VIK}, purely kinetic multi k-essence model crossing the phantom divide barrier \cite{CHIR2}, \cite{SSDAS}, tracking solution \cite{DKS} and so on.

Some authors have shown that there is  an algorithm for constructing potential associated to solutions 
on a FRW brane with a power law behavior when the matter fields are a tachyon  as well as  a scalar field  \cite{PAD1}, \cite{SA1}. The high energy limit for the tachyonic potential on the brane turns out to be $V= V_{0}\phi^{-1}$  and reduces to inverse square potential in the low energy limit. Our goal in this brief article is to  show that the quadractic brane correction shifts the inverse square potential, characterizing the FRW cosmology, to the inverse linear one at high energy.  Another complentary motivation for our analysis is to classify the different kinds of singularities, in particluar, those related with an accelerated or super-accelerated regime. 

This letter is organized as follows. Our starting point is to present the model of k-essence matter localized on the FRW brane, hereafter  we obtain the scale factor and potential for a k-essence field evolving linearly with time when the speed of sound is not infinite. After that we examine the divergent sound speed model. Then, we apply these results to the extended tachyon field case. Finally the conclusions are stated. We will take the gravitational coupling constant $8\pi G = 1$ throughout this article.

%%%%%%%%%%%%%%%%%%%%%%%%%%%%%%%%%%%%%%%%%%%%%%%%%%%%%%%%%%%%%%%%%%%%%%%%%%%%%%%%%%%%%%%%%%%%%%%%%%%%%%%%%%%%%%%%%%%%%%%%%%%%%%%%%%%%%%%%
\section{k-essence in brane-world model}
%%%%%%%%%%%%%%%%%%%%%%%%%%%%%%%%%%%%%%%%%%%%%%%%%%%%%%%%%%%%%%%%%%%%%%%%%%%%%%%%%%%%%%%%%%%%%%%%%%%%%%%%%%%%%%%%%%%%%%%%%%%%%%%%%%%%%%%%%%
In the following, we shall explore the evolution of a cosmological brane filled with a k-essence field $\phi$. The energy-momentum tensor of the k-field on the brane is

\be
\n{t}
T_{\mu\nu}=V(\phi)\left[2F_{,X}\nabla_{\mu}\phi\nabla_{\nu}\phi-g_{\mu\nu}F\right],
\qquad F_{,X}=\frac{d\,F}{d\,X},
\ee
where $g_{\mu\nu}$ is the four dimensional metric with signature $(-,+,+,+)$, $F(X)$ is a function of the kinetic energy $X\equiv \nabla_\nu \phi\nabla^{\nu}\phi=-\dot{\phi}^{2}$ and $V(\phi)$ is a positive  potential.  Using the perfect fluid analogy, the energy density and pressure are given by 

\be
\n{ro}
\rho_{\phi}=V(\phi)[F-2XF_{,X}], \qquad  p_{\phi}={\cal{ L}}=-V(\phi)F(X).
\ee
We will focus on cosmological branes  with a flat FRW spacetime as the induced metric 
\begin{equation}
ds^{2}=-dt^{2}+ a^{2}(t)(dx^{2}+dy^{2}+dz^{2}),
\end{equation}
where $a(t)$ is the scale factor and  $H=\dot{a}/a$ is the Hubble expansion rate. Then the modified Einstein equations on the brane are (see \cite{BRA1}, \cite{BRA2} and \cite{review})
 \ben
 3H^{2}=\rho_{\phi}+ \frac{3}{\lambda^{2}}\rho_{\phi}^2 \label{fm}
,\\
{\dot\rho}_{\phi}+3H(\rho_{\phi}+p_{\phi})=0.\label{cphi1}
 \een
The evolution of the early universe can be separated into the high  energy regime, $\rho_{\phi}>>\lambda^2$, where the quadratic term in  Eq.
(\ref{fm}) dominates getting an unconventional equation $H^2\approx\rho_{\phi}^{2}/\la^{2}$ and the low energy regime, $\rho_{\phi}\ll\lambda^{2}$, where the linear density term dominates recovering the Friedmann equation in four dimension $3H^{2}\approx\rho_{\phi}$. 

As consequence of Eqs. (\ref{ro}) and (\ref{cphi1})  the dynamic equation for the $k$-essence field can be recast as
\be
\n{kg}
[F_{,X}+2XF_{,XX}]\ddot\phi+3HF_{,X}\dot\phi+\frac{V'}{2V}[F-2XF_{,X}]=0,
\ee
with $'\equiv d/d\phi$. Also, by writing the equation of state  for the k-essence as $p_{\phi}=(\gamma_{\phi}-1)\rho_{\phi}$, the barotropic index reads $\gamma_{\phi}=-2XF_{,X}/(F-2XF_{,X})$.
%\be
%\label{gamma}
%\gamma_{\phi}=\frac{-2xF_{x}}{F-2xF_{x}}.
%\ee
The simplest solution of Eqs. (\ref{fm}) and (\ref{cphi1}) is obtained for $X=X_0$, with $X_0$ a constant and $F_{,X}(X_0)=0$. For these kinetic functions the barotropic index vanishes  so using  Eq. (\ref{cphi1}) one gets a constant energy density [$\dot\ro_\phi=0$]. It means that the potential is a constant in time also, then the Eq. (\ref{fm}) constraints the expansion rate to be constant also and the brane exhibits a de Sitter phase for those $F$. 

Other kind of solutions can be found by re-writing the modified Friedmann equations on brane (\ref{fm})-(\ref{cphi1}) as follows
\ben
 &&3H^{2}=V{\al}+ \frac{3}{\lambda^2}{\al}^2V^{2}\label{fm2}
,\\
&&\frac{\dot{\al}}{{\al}}+\frac{\dot{V}}{V} +3H\gamma_{\phi}=0,\label{cphi2}
 \een
where we have introduced a new variable ${\al}=\rho_{\phi}/V=F-2XF_{,X}$ for simplicity. This new setting is useful because it tells us how to obtain exact solutions with k-essence source in a neatly way if the kinetic function $F(X)$ fulfills some simple criteria. Now, in order to carry on we need to  assume two simple conditions: (i)- a k-field evolving linearly with the cosmological time, $\phi=\phi_{0}t$ with $\phi_{0}$ a constant, and (ii)- the following constraint ${\al}={\al}_{0} $ is fulfilled. 

One would like to describe the solutions in which the effective speed sound does not vanish [$F_{,X}/c^{2}_{s}=2XF_{,XX}+F_{,X}\neq \infty$], such that the term proportional to $c^{2}_{s}\ddot{\phi}$ in Eq.$\ref{kg}$ is well defined when the field $\phi$ evolves linearly with the cosmic time. Taking ${\al}={\al}_{0}$ along with $\phi=\phi_{0}t$ one can assure that the conditions $X_{0}=-\phi^{2}_{0}$, $F(X_{0})=\mbox{const.}$, and $F_{,X}(X_{0})=\mbox{const.}$ are satisfied. With the latter ansatz it turns to be that the barotropic index becomes constant $\ga_\phi=\ga_0$ so one can easily integrate Eq.(\ref{cphi2}) for getting  as result the potential in terms of the scale factor: $V=V_{0}a^{-3\gamma_{0}}$. For this type of potential  one finds that the Eq. (\ref{fm2}) reads
\be
\label{fmadet}
{\dot{y}}^{2}=3\ga_0^2\left[\ro_{0}y +\frac{3\ro^{2}_{0}}{\la^{2}}\right]
\ee
with $y=a^{3\gamma_{0}}$, $\ro_{0}=V_{0}{\al}_{0}$ and its general solution is given by 
\be
\label{adet}
a(\tau)=(3\ro_0)^{1/3\gamma_{0}}\left[ \frac{\tau^2}{4}\pm\frac{\tau}{\la}\right]^{1/3\gamma_{0}}
\ee
where $\tau=\gamma_{0}t$. This solution has singularities at $\tau_s=0$ and/or  $\tau_s=\mp 4/\la$. For $\ro_0>0$, we have four expanding universes, two of them with $\gamma_{0}>0$ evolve from an initial singularity at $\tau_s=0$ or $\tau_s=4/\la$.  The scale factor begins as $a\propto (\tau-\tau_s)^{1/3\gamma_{0}}$ in the high energy regime and ends as $a\propto \tau^{2/3\gamma_{0}}$ in the low energy regime. The remaining two universes with $\ga_0<0$ end in a final big rip at $\tau_s=0$ or $\tau_s=-4/\la$ having the final behavior $(\tau_s-\tau)^{1/3\ga_0}$. Also, we have four contracting universes by making the time reversal of the above solutions. For $\ro_0<0$, the solutions have an extremum at $\tau_{e}=2/\la$ where the scale factor takes the value $a_e=(-3\ro_0/\la^2)^{1/3\ga_0}$. They represent two universes with a finite time span, one of them begins from an initial singularity reaches a maximum value $a_e$ and ends in a big crunch at $\tau_s=4/\la$. The other begins from a singularity at $\tau_s=0$, reaches a minimum at $a_e$, where it bounces, and ends in a final big rip at $\tau_s=4/\la$. In addition we have the time reversal of these solutions also. It is important to remark that some authors have studied accelerated or decelerated scenarios of 3-brane \cite{ADBrane} and also bouncing brane \cite{BounceBrane}. Interestingly,  a DGP cosmology filled with k-essence field  was studied in \cite{MBLL}, obtaining that  the asymptotic behavior of the brane can involve accelerating regime, big bangs, big crunches, big rips or quiescent singularities.

On the other hand, combining $\phi=\phi_{0}t$ with $a(t)$ and $V=V_{0}a^{-3\gamma_{0}}$, we obtain  the following potential 
\begin{equation}
\label{Vphi}
V(\phi)=\frac{V_0}{3\ro_0}\left[\frac{\ga_0^2 \phi^2}{4\phi_0^2}\pm\frac{\ga_0\phi}{\la\phi_0}\right]^{-1}.
\end{equation}
In the early universe where the brane cosmology takes place, the strong  effects on the brane become very important i.e., the Hubble parameter behaves as $H\propto \rho/\la$ when $\la\rightarrow 0$ and the potential approaches $V\propto \pm\la \phi^{-1}$. It seems to be the counterpart of the  solution found in the context of  cosmological brane filled with  quintessence field which is driven by an inverse square potential \cite{PD}, \cite{KIEM}, \cite{NOS}. In the opposite limit $\la\rightarrow \infty$, the potential takes the form $V\propto \phi^{-2}$ and we recover the power law solution for the scale factor with an inverse square potential \cite{PAD1}, \cite{AFE1} \cite{CHFE1}. 

The potential (\ref{Vphi}) is interesting because it  shows that the brane correction shifts the inverse square power law, characterizing the Friedmann cosmology, to the inverse linear one at high energy. It has a global minimum at $\phi_{e}=2\phi_0/\la\ga_0$ or a global maximum at $\phi_{e}=-2\phi_0/\la\ga_0$. 

When one relates the dynamical equations for the 3-brane  with the standard Friedmann framework throught an effective pressure and an effective energy density, then one finds a linear barotropic equation of state in the low energy limit. However, in the high energy  regime the k-essence source turns out to be a modified Chaplygin gas, $P_{ef}\approx A\ro_{ef}+B\ro^{-\alpha}_{ef}$ with $A=(2\gamma_{0}-1)$, $B=\pm\gamma_{0}\la/\sqrt{3}$ and $\alpha=-1/2$.

Now we are going to study the group of solutions related to the atypical k-essence model  \cite{CHI1}, \cite{CHFE1}, \cite{CHIR1}  where its kinetic funcion $F^\infty$ satisfies the requisites $\al=\al_0$ and $F^\infty-2XF^\infty_{,X}={\al}_{0}$. Integrating the latter condition one finds the kinetic function of this theory,
\begin{equation} 
\label{fdiv}
F^\infty={\al}_{0}  + \beta\sqrt{-X}
\end{equation}
where ${\al}_{0}$ and $\beta$ are arbitrary constants. In this case  we work under the condition $F_{,X}+2XF_{,XX} = 0$, i.e., the solutions will be  associated with a divergent sound velocity \cite{MCL}, \cite{CHIR1} as well as  the extended tachyon model considered in \cite{CHI1}. In Ref. \cite{CHIR1} it was shown that the existence of solutions for the Friedmann cosmology  leads to the inverse square potential.

The Hubble parameter can be obtained
by substituting Eq. (\ref{fdiv}) and the barotropic index [$\gamma_{\phi}=-\beta\sqrt{-x}/\al_0$], associated with the particular kinetic function (\ref{fdiv}),  into Eq. (\ref{cphi2})
\begin{equation}
\label{hv}
H=\frac{\al_0 V'}{3\beta V}.
\end{equation}
This atypical model has an energy density $\rho_{a}=\al_0 V(\phi)$, so replacing (\ref{hv}) into the Eq. (\ref{fm}) one gets the potential
\begin{equation}
\label{Vaty}
V(\phi)=\frac{1}{3}\left[\frac{\beta}{4\al_0}\phi^{2}\pm \frac{\phi\beta}{\la}\right]^{-1}.
\end{equation}
Although potentials (\ref{Vphi}) and (\ref{Vaty}) have the same form, the parameters have a different meaning. The potential on the brane (\ref{Vaty}) generalizes the inverse square one of the Friedmann cosmology shifting the divergence to $\phi_{\infty}=\pm4\al_0/\la\beta$. There is no equation of evolution for $\phi$ nor $a$ in the sense that we need first to know one of them to get the other one. Hence, the time dependence of $a$ or $\phi$ is not fixed by the form of the potential, meaning that, the modified Friedmann equation on the brane has some missing information when the atypical k-essence is considered. In other words the dynamic of this theory is uncomplete. This atypical k-essence theory becomes an intrinsic component of all k-essence models. Therefore, for asymptotic power-law expansions the linear k-field model driven by a nearly inverse square potential and the atypical model are isomorphic.

%%%%%%%%%%%%%%%%%%%%%%%%%%%%%%%%%%%%%%%%%%%%%%%%%%%%%%%%%%%%%%%%%%%%%%%%%%%%%%%%%%%%%%%%%%%%%%%%%%%%%%%%%%%%%%%%%%
\subsection{Extended Tachyons in brane-worlds}
%%%%%%%%%%%%%%%%%%%%%%%%%%%%%%%%%%%%%%%%%%%%%%%%%%%%%%%%%%%%%%%%%%%%%%%%%%%%%%%%%%%%%%%%%%%%%%%%%%%%%%%%%%

It has been recently proposed  an extended kinetic function, which contains the tachyon kinetic function as a special case, and which at the same time it  allows the associated barotropic index takes any value \cite{CHI1}. Refs. \cite{CHI1} and \cite{CFKR} provide an adequate framework to consider phantom tachyon as well as complementary tachyon in addition to the ordinary one. This one new species of tachyons generalize the Chaplygin gas on having introduced the phantom  as well  complementary Chaplygin gases. The kinetic functions for three above-mentioned cases are
\begin{eqnarray}
\n{to}
F_{t}=(1-\dot{\phi}^{2r}_{t})^{1/2r},   \qquad   0<\ga_{t}=\dot{\phi}^{2r}_{t}<1 
\\
\n{tp}
F_{ph}=(1+\dot{\phi}^{2r}_{ph})^{1/2r},   \quad   -\infty<\ga_{ph}=-\dot{\phi}^{2r}_{ph}<0 
\\
\n{ft}
F_{c}=-(\dot{\phi}^{2r}_{c}-1)^{1/2r},   \quad   1<\ga_{c}=\dot{\phi}^{2r}_{c}<\infty
\end{eqnarray}
where $r$ is a real parameter. 

The scale factor and potential for the three sets of extended tachyon fields are given by Eqs. (\ref{adet}) and (\ref{Vphi}) with $\gamma_{0}=\phi^{2r}_{0t}$, $\gamma_{0}=-\phi^{2r}_{0ph}$, $\gamma_{0}=\phi^{2r}_{0c}$ and ${\al}_{0}=(1-\phi^{2r}_{0t})^{-1/2}$, ${\al}_{0}=(1+\phi^{2r}_{0ph})^{-1/2}$, ${\al}_{0}=-(\phi^{2r}_{0c}-1)^{-1/2}$ respectively. 

The ordinary and complementary tachyons lead to expanding scenarios. However, the former has a final accelerated expansion for $\phi^{2r}_{0t}<2/3$. On the other hand, a universe dominated by a phantom tachyon field ends in a final big rip at $\tau_s=0$ or $\tau_s=-4/\la$ where the scale factor blows up as $a\propto \tau^{-1/3\phi^{2r}_{0ph}}$  and  the potential diverges as $V\propto \phi^{-1}_{ph}$. Also, we comment that the  ordinary tachyon field as well as the complementary one satisfy the weak energy condition, namely $\ro\geq0$ and $\ro+p\geq0$, and the null energy condition, that is $\ro+p\geq0$. However, as one would expect the phantom tachyon field violates both conditions. Further, late accelerated phase with tachyon fields was examined in \cite{LateTachyonAcc} and  scaling solutions in  modified gravity sourced by tachyons were found in \cite{ScalingTachyonBrane}.

An interesting possibility arises for the extended tachyon matter localized on the brane. Taking into account that the modified Friedmann equation are quadratic in the local energy density, we consider a possible tachyon field with negative energy density. In order to do so, we change the sign of the kinetic function in Eqs. (\ref{to})-(\ref{ft}) keeping the potential positive definite $V>0$. Note that the barotropic index preserves its sign under the change $F\to -F$.

A universe filled with these extended tachyons evolves between two singularities having a finite time span $t_{ts}=4/\la|\gamma_{0}|$. The scale factor has a maximum $a=a_e$ for ordinary and complementary tachyons, ending in a big crunch or has a minimum $a=a_e$ where the phantom tachyon bounces before  it blows up in a big rip. A universe filled with complementary tachyon collapses faster than the one with ordinary tachyon while the potential has a global minimun at $a=a_e$ or a global maximum at $a_e$ for an universe filled with phantom tachyon. 

As a final remark, if we choice $\ga_0=1$, which means to select the matter energy density  $\rho=\rho_0 a^{-3}$, then the Friedamn equation on the brane world model with zero nonlocal radiation term contains the same terms than the Friedmann equation for a Bianchi Type-I universe matter dominated. Basically, the quadratic contribution introduced by the brane behaves as stiff matter, the same as the contribution of the anisotropy of the universe at early stage contained in the shear term $\sigma=\sigma_{0}/a^{6}$.
This analogy seems reasonable because the anisotropy as well as the brane effects exhibit their main features at early stage of the universe. 

\section{Conclusions}
 We have explored a 3-dimensional cosmological brane filled with  a k-essence field for spatially flat FRW spacetime. We have described the dynamical evolution of brane when the k-field evolves linearly with the cosmic time, obtaining as  result that scale factor has one or two singularities depending on the sign of $\rho_{0}$ and $\gamma_{0}$. Later on we have classified the different solutions. This classification includes expanding or contracting brane-worlds for $\ro_{0}>0$ with a power law behavior at the initial and final stages while for $\ro_{0}<0$ there are  bouncing universes with a finite time span and a final big cruch for $\ga_0>0$ or a final big rip for $\ga_0<0$. The linear k-field is driven by an inverse quadratic polynomial potential. So that, the quadratic brane correction shifts the inverse square potential characterizing the Friedmann cosmology, to the inverse linear one at hight energy where the k-essence  is represented by a modified Chaplygin gas.
We have obtained a similar potential by taking into account the atypical k-essence model and showed that the dynamic evolution of $\phi$ and $a$ is not fixed neither in the brane scheme nor at Friedmann cosmology, indicating the uncomplete realization of this model.
Finally, we have introduced the extended tachyons with negative energy density. Within the brane context the scale factor evolves between two singularities having a finite time span which depends on the brane tension and the barotropic index. For the ordinary and complementary negative tachyons the scale factor has a final big cruch  while for the phantom one bounces and ends in a final big rip.

%%%%%%%%%%%%%%%%%%%%%%%%%%%%%%%%%%%%%%%%%%%%%%%%%%
\acknowledgments
%%%%%%%%%%%%%%%%%%%%%%%%%%%%%%%%%%%%%%%%%%%%%%%%%
LPC thanks the University of Buenos Aires for partial support under
project X044, and the Consejo Nacional de Investigaciones
Cient\'{\i}ficas y T\'ecnicas under project 5169.  MGR is supported by  the Consejo Nacional de Investigaciones Cient\'{\i}ficas y T\'ecnicas (CONICET).

%%%%%%%%%%%%%%%%%%%%%%%%%%%%%%%%%%%%%%%%%%%%%%%%%%%%%%%%%%%%%%%%%%%%%%%%%%%%%%%%%%%%%%%%%%%%%%%%%%%%%%%

\end{document}